\documentclass[12pt,preprint]{aastex}

\shorttitle{Nonlinear Magnetothermal Instability}
\shortauthors{Parrish et al.}

\begin{document}

\title{Nonlinear Evolution of the Magnetothermal Instability in Two Dimensions}

\author{Ian J. Parrish and James M. Stone\altaffilmark{1}}
\affil{Department of Astrophysical Sciences, Princeton University, Princeton, NJ 08544}

\altaffiltext{1}{Program in Applied and Computational Mathematics, Princeton University, Princeton, NJ, 08544}

\begin{abstract}

In weakly magnetized, dilute plasmas in which thermal conduction along
magnetic field lines is important, the usual convective stability
criterion is modified.  Instead of depending on entropy gradients,
instability occurs for small wavenumbers when $(\partial P/\partial
z)(\partial \ln T / \partial z) > 0$, which we refer to as the Balbus
criterion.  We refer to the convective instability that results in this
regime as the magnetothermal instability (MTI).  We use numerical
MHD simulations which include anisotropic electron heat conduction to
follow the growth and saturation of the MTI in two-dimensional, plane
parallel atmospheres that are unstable according to the Balbus criterion.
The linear growth rates measured in the simulations agree with the weak
field dispersion relation.  We investigate the effect of strong fields
and isotropic conduction on the linear properties and nonlinear regime
of the MTI.  In the
nonlinear regime, the instability saturates and convection decays away,
when the atmosphere becomes isothermal.  Sustained convective turbulence
can be driven if there is a fixed temperature difference between the top
and bottom edges of the simulation domain, and if isotropic conduction
is used to create convectively stable layers that prevent the formation
of unresolved, thermal boundary layers.  The
largest component of the time-averaged heat flux is due to advective motions. 
These results have implications for a variety of astrophysical systems,
such as the temperature profile of hot gas in galaxy clusters,
and the structure of radiatively inefficient accretion flows.

\end{abstract}


\section{Introduction} \label{introduction}

In dilute astrophysical plasmas, the mean free path between particle
collisions can be much larger than the ion gyroradius.  In this
circumstance, the equations of magnetohydrodynamics (MHD) which
describe the dynamics of the plasma must include anisotropic transport
terms for energy and momentum due to near free-streaming motions of
particles along magnetic field lines \citep{brag65}.  The parallel
thermal conductivity of the electrons is larger than that of the ions by a
factor proportional to $(m_i/m_e)^{1/2}$, whereas the parallel viscosity
of the ions is larger than that of the electrons by the same factor.
Thus, provided that collisions are still frequent enough to keep the
distributions of the velocity components parallel and perpendicular to
the magnetic field in equilibrium, the usual equations of MHD must be
supplemented with an anistropic electron heat conduction term, and an
anisotropic ion viscosity.  Moreover, since the ratio of ion viscosity
to electron heat conduction is proportional to $(m_e/m_i)^{1/2}$, it is
often sufficient to neglect the former, and consider only the effect of
anisotropic electron heat conduction.  In a rotating system, even a small ion viscosity may be important \citep{bal04}. If the parallel and perpendicular
temperature of particles are not in equilibrium, more complex closures
are required \citep{hp90}.

The implications of anisotropic transport terms on the overall dynamics
of dilute astrophysical plasmas is only beginning to be explored
\citep{bal01, qdh02, sha03}.  One of the most remarkable results obtained thus
far is that the convective stability criterion for a weakly magnetized
dilute plasma in which anisotropic electron heat conduction occurs is
drastically modified from the usual Schwarzschild criteria \citep{bal00}.
In particular, stratified atmospheres are unstable if they contain a
{\em temperature} (as opposed to 
{\em entropy}) profile which is decreasing upward.  There are intriguing
analogies between the stability properties of rotationally supported
flows (where a weak magnetic field changes the stability criterion from a
gradient of specific entropy to a gradient of angular velocity), and the
convective stability of stratified atmospheres (where a weak magnetic
field changes the stability criterion from a gradient of entropy to a
gradient of temperature).  The former is a result of the magnetorotational
instability \citep[MRI;][]{bh98}.  The latter is a result of anisotropic
heat conduction.  To emphasize the analogy, we will refer to  this new
form of convective instability as the magnetothermal instability (MTI).
The MTI may have profound implications for the strucuture and dynamics
of many astrophysical systems.

In this paper, we use numerical methods to explore the nonlinear
evolution and saturation of the MTI in two-dimensions.  We adopt
an arbitrary vertical profile for a stratified atmosphere in which
the entropy increases upward (and therefore is stable according to
the Schwarzschild criterion), but in which the temperature is decreasing
upwards (and therefore is unstable according to the Balbus criterion, $(\partial P/\partial
z)(\partial \ln T / \partial z) > 0$).
We confirm the linear
growth rates predicted by \citet{bal00} for dynamically weak magnetic
fields, and numerically measure the growth rates for stronger fields.
We follow the evolution of the instability into the nonlinear regime,
and show that it results in vigorous convective turbulence and heat transport.
These results may have implications for radially stratified atmospheres
in which anisotropic transport may be present, including x-ray emitting
gas in clusters of galaxies \citep{fab94, zak03}, atmospheres of strongly
magnetized neutron stars, and radiatively inefficient accretion flows
\citep{spb99, nmq98}.

This paper is organized as follows.  In \S2, we describe our numerical
methods and initial conditions.  In \S3, we compare the computational
results to the analytical results of linear theory. In \S4 and \S5 we present
the results from the non-linear regime and saturated states for two
different choices of boundary conditions for the temperature at the top
and bottom of the computational domain.  Finally, in \S6 we summarize our
results, discuss applications, and describe future work.

\section{Method} \label{method}

\subsection{Equations of MHD with Anisotropic Heat Conduction}

The equations of MHD with the addition of the heat flux, $\mathbf{Q}$,
and a vertical gravitational acceleration, ${\bf g} = -g_{0}{\bf \hat{z}}$ are
\begin{equation}
\frac{\partial \rho}{\partial t} + \mathbf{\nabla}\cdot\left(\rho \mathrm{\mathbf{v}}\right) = 0,
\label{MHD_continuity}
\end{equation}
\begin{equation}
\frac{\partial(\rho\mathbf{v})}{\partial t} + \mathbf{\nabla}\cdot\left[\rho\mathbf{vv}+\left(p+\frac{B^{2}}{8\pi}\right)\mathbf{I} -\frac{\mathbf{BB}}{4\pi}\right] + \rho\mathbf{g}=0,
\label{MHD_momentum}
\end{equation}
\begin{equation}
\frac{\partial\mathbf{B}}{\partial t} + \mathbf{\nabla}\times\left(\mathbf{v}\times\mathbf{B}\right)=0,
\label{MHD_induction}
\end{equation}
\begin{equation}
\frac{\partial E}{\partial t} + \mathbf{\nabla}\cdot\left[\mathbf{v}\left(E+p+\frac{B^{2}}{8\pi}\right) - \frac{\mathbf{B}\left(\mathbf{B}\cdot\mathbf{v}\right)}{4\pi}\right] 
+\mathbf{\nabla}\cdot\mathbf{Q} +\rho\mathbf{g}\cdot\mathbf{v}=0,
\label{MHD_energy}
\end{equation}
where the symbols have their usual meaning. The total energy $E$ is given as
\begin{equation}
E=\epsilon+\rho\frac{\mathbf{v}\cdot\mathbf{v}}{2} + \frac{\mathbf{B}\cdot\mathbf{B}}{8\pi},
\label{MHD_Edef}
\end{equation}
with the internal energy, $\epsilon=p/(\gamma-1)$.  For this paper, we assume
$\gamma=5/3$ throughout.

The heat flux contains contributions both from electron motions (which are
constrained to move primarily along field lines) and from isotropic transport
which may arise due to photons or particle collisions which drive cross
field diffusion.  Thus, 
${\bf Q} = {\bf Q}_{C} + {\bf Q}_{R}$, where 
\begin{equation}
\mathbf{Q}_{C} = - \chi_{C} \mathbf{\hat{b}\hat{b}}\cdot\mathbf{\nabla}T,
\label{coulombic}
\end{equation}
\begin{equation}
\mathbf{Q}_{R} = - \chi_{R} \mathbf{\nabla}T,
\label{radiative}
\end{equation}
where $ \chi_{C} $ is the Spitzer Coulombic conductivity \citep{spitz62},
$\mathbf{\hat{b}}$ is a unit vector in the direction of the magnetic field,
and $ \chi_{R} $ is the coefficient of isotropic conductivity, ostensibly due to radiation.
We consider both $ \chi_{C} $
and $ \chi_{R} $
as free parameters in the problem and will vary both independently.

\subsection{Initial Equilibrium Conditions}

We now specify a vertical equilibrium state in which we can study the
linear modes and nonlinear evolution of the MTI.  The first step is
to choose an \textit{Ansatz} for the form of the gravitational field.
In this paper, we adopt the simplest choice of a constant vertical
acceleration: $g(z) = -g_{0}$.  This form is appropriate for a
thin plane-parallel atmosphere.  Alternative choices in which the
gravitational acceleration varies with height would be appropriate for
the vertical structure of a thin accretion disk, or the radial structure
of a self-gravitating gas sphere.  However, since the MTI is a local
instability, the evolution of modes with wavelengths much smaller than
the scale height, $H \sim c_{s}^{2}/g$, where $c_{s}^2$ is the adiabatic sound speed,
 should be independent of
the form of $g(z)$.  We will consider other profiles in applications
specific to particular astrophysical systems in future work.

We next make the \textit{Ansatz} that the temperature in the atmosphere
decreases with height, while the entropy increases with height.  Perhaps the
simplest vertical equilibrium state which satisfies this condition is a
power law:
\begin{equation}
T(z) = T_{0}(1-z/z_{0}),
\label{T_profile}
\end{equation}
\begin{equation}
\rho(z) = \rho_{0}(1-z/z_{0})^{2},
\label{rho_profile}
\end{equation}
\begin{equation}
p(z) = p_0(1-z/z_{0})^{3},
\label{P_profile}
\end{equation}
where $z_{0}$ is a constant.  We have assumed that the magnetic pressure
is small compared to the gas pressure ($\beta = 8\pi p/B^{2} \gg 1$),
so that the vertical structure is given by the solution of the equation
of hydrostatic equilibrium.  Of course, any power-law profile with $T$
decreasing with height would do; however, by choosing a linear profile
we guarantee that the vertical gradients are shallow and easier to resolve
numerically.  To satisfy hydrodynamic stability, we choose $T_{0}=\rho_{0}=p_{0}=g_{0}=1$ in an appropriate system of units. Convective stability places a constraint on the entropy
gradient via the Schwarzschild criterion, namely, $dS / dz > 0$.  For our
choice of initial conditions convective stability to perturbations requires, $0 < z_{0}^{-1} < 2/5 g_{0}$.
Setting $z_{0}=3$ satisfies the hydrodynamic equilibrium and convective stability constraints. These constants determine the adiabatic sound
speed in the domain to be
\begin{equation}
c_{s}^{2}(z)=\frac{\gamma p(z)}{\rho(z)} = \frac{5}{3}\left(1-\frac{z}{3}\right).
\label{soundspeed}
\end{equation}
By using a simulation domain which is small compared to the scale height,
$H=C_{s}^{2}/g \approx 1.67$, we ensure the sound speed does not vary much 
from the top to the bottom of the box.

Finally, we add a weak horizontal magnetic field so that the domain contains either a zero or net magnetic flux:
\begin{equation}
B_{x}(z) = \left\{ \begin{array}{ll}
B_{0} & \textrm{net flux} \\
B_{0}\sin\left(\frac{2\pi z}{L_{z}}\right) & \textrm{zero flux;}
\end{array} \right.
\label{B_profile}
\end{equation}
The Alfv\'{e}n speed becomes
\begin{equation}
v_{A}^{2}(z)=\frac{B_{x}^{2}(z)}{4\pi\rho(z)}=\frac{B_{0}^{2}}{4\pi}
\left(1-\frac{z}{3}\right)^{-2}.
\label{alfvenspeed}
\end{equation}
for the case of a uniform horizontal field.  The value of $B_{0}$ is
chosen to be small so that tension effects are unimportant in the linear
regime, that is typically $v_{A} \ll c_{s}$.  We also investigate the
effects of strong fields in suppressing the instability.

\subsection{Linear Stability Properties of the MTI}

If $\chi_{R}=0$, the heat flux in the initial state is zero because the
magnetic field lines are parallel to the isotherms, and it represents an
equilibrium.  Now imagine a small vertical perturbation that partially
aligns the magnetic field with the background temperature gradient,
that is $\mathbf{B} \cdot \bf{\nabla} T \neq 0$.  Heat is now able
to flow from the hotter to the colder regions, causing them to become
buoyant and rise.  This, in turn, causes the magnetic field to become
more aligned with the background temperature gradient; increasing the
heat flow.  Thus, a growing instability is generated.

A quantitative understanding of the instability is gained by considering
the linearized equations of MHD (eq.[\ref{MHD_continuity}-\ref{MHD_Edef}])
with the specification of a heat flux that is purely Coulombic.
The details can be found in \citep[section 4,][]{bal00}. 

We introduce the Brunt-V\"ais\"al\"a frequency, $N$,
\begin{equation}
N^{2} = -\frac{1}{\gamma \rho}\frac{\partial P}{\partial z}\frac{\partial \ln S}{\partial z},
\label{brunt-vaisala}
\end{equation}
the frequency of vertical oscillations in a stably
stratified atmosphere.  Preserving Balbus' notation, we also define two useful
quantities,
\begin{equation}
\chi_{c}' = \frac{\gamma - 1}{P}\chi_{c} \qquad \textrm{and} \qquad \chi' = \frac{\gamma - 1}{P}\left(\chi_{c}+\chi_{R}\right).
\label{cond_normalization}
\end{equation}
Using the Fourier convention for perturbations, $\exp{(\sigma t + ikx)}$,
and in the limit of a weak magnetic field, $k^{2}v_{A}^{2} \ll 1$,
the dispersion relation simplifies to the non-dimensionalized form
\begin{equation}
\left(\frac{\sigma}{N}\right)^{3} + \frac{1}{\gamma}\left(\frac{\sigma}{N}\right)^{2}\left(\frac{\chi' T k^2}{N}\right) +\left(\frac{\sigma}{N}\right) + \frac{d \ln T}{d \ln S}\left(\frac{\chi'_{c} T k^2}{N}\right) = 0,
\label{dispersionrelation}
\end{equation}

Solutions to the dispersion relation are plotted in Figure 1. 
The lower branches are stable oscillations.  Note that in Figure 5
of (Balbus 2000) the factor of $1/\gamma$ multiplying the second term
of equation \ref{dispersionrelation} is omitted.  Our choice of initial
conditions sets the constant $d\ln T/d \ln S = -3$.  By analysis of the
Routh-Hurwitz criterion \citep[Section 4.3,][]{bal00}, the instability
criterion is
\begin{equation}
k^2v_{A}^{2} - \frac{\chi'_{C}}{\rho \chi'}\frac{\partial P}{\partial z}\frac{\partial \ln T}{\partial z} < 0.
\label{MTI_instability_criterion1}
\end{equation}
In the limit of infinitesimal wavenumber, the instability criterion
simply requires the temperature and pressure gradients to be in the same
direction, \textit{i.e.},
\begin{equation}
\frac{\partial P}{\partial z}\frac{\partial \ln T}{\partial z} > 0.
\label{MTI_instability_criterion2}
\end{equation}
We will refer to the instability criterion equation (18) as the Balbus
criterion.
The instability criterion of the magnetorotational instability
\citep{bh98} can be written
\begin{equation}
k^2v_{A}^2 + \frac{d\Omega}{d\ln R} < 0,
\label{MRI_instability_criterion}
\end{equation}
where $\Omega$ is the angular velocity.  The similarity between the MRI
and the MTI is self-evident.  Strong magnetic fields will stabilize
short wavelength perturbations in both instabilities through tension.
This point will be addressed in \S\ref{linearmag} in more detail.

\subsection{Numerical Algorithms for studying the Nonlinear Regime}
\label{compmethods}

We follow the MTI into the nonlinear regime using the Athena MHD
code \citep{gs05}.  Athena is an unsplit Godunov method for ideal MHD
that utilizes constrained transport to preserve the divergence-free
constraint of the magnetic field.
Isotropic and anisotropic heat conduction are added to the basic MHD
algorithm through operator splitting.  By finite differencing the
diffusion terms in local field line coordinates, we are able to keep
the ratio of parallel to perpendicular thermal diffusion greater than at
least $10^3$ for any orientation of the field with respect to the grid
(and much larger than that when the field is closely aligned with grid).
We show in section 3.3 that the level of isotropic conduction added
by numerical truncation error has a negligible effect on the linear
properties of the MTI.  Details of the numerical methods used for
anisotropic conduction, including tests, are given in the appendix.

All of the simulations presented in this paper are performed on a two
dimensional grid, with horizontal and vertical coordinates $x$ and $z$.
The size of the domain is $L_x = L_z = 0.1$.  Our standard numerical
resolution is $100^{2}$ for single-mode studies and $128^{2}$ for multi-mode studies, although we have performed resolution studies
which span grid resolutions from $64^{2}$ to $512^{2}$.  By using a box
size which is small compared to the scale height, we ensure the sound
speed varies only slightly from the top to the bottom of the domain.  From
equation (11) the average sound speed is $\bar{c_{s}}\approx 1.28$, giving
a sound crossing time, $\bar{\tau_{s}} \approx 7.81 \times 10^{-2}$.
For comparison, the typical magnetic field strength in the initial state
is chosen to be $B_0/\sqrt{4\pi} = 10^{-3}$, giving an Alfv\'{e}n speed of
$v_{A} = 10^{-3}$, and an Alfv\'{e}n crossing time of 100.

We use periodic boundary conditions in the horizontal $x$-coordinate.
We study the saturation of the MTI using two different boundary conditions
in the vertical $z$-coordinate.  In both cases, reflecting boundary
conditions are applied to all variables except the internal energy, in
which an extrapolation is used so that the pressure gradient balances the
gravitational acceleration at the boundary---an important consideration for Godunov schemes (see below).  In addition, for the first
case, the boundary temperature is set equal to the temperature in the
last active computational cell in the grid; that is a Neumann boundary
condition is used for the temperature.  We will refer to this boundary
condition for the temperature as ``adiabatic boundaries,'' as heat is
not exchanged.  For the second case, the  boundary temperature is held
fixed at constant values at the upper and lower boundaries; that is a
Dirichlet boundary condition is used for the temperature.  This latter
case drives a heat flux into and out of the domain.  We will refer to
this boundary condition for the temperature as ``conducting boundaries.''

By periodicity, there can be no net flux of mass, momentum, or energy in
the horizontal direction.  However, because we do not enforce $V_{z}=0$
on the vertical boundaries explicitly, it is possible through numerical
truncation error to have a net flux of mass or momentum through
the vertical boundaries.  Note we specifically permit a net flux of energy (heat) through the vertical boundary.   For Godunov schemes like Athena, in order for
the flux difference due to the vertical pressure gradient to exactly
balance the gravitational acceleration source term, the gravitational source terms
must be incorporated directly into the computation of the fluxes.  This requires substantial modification of the underlying MHD algorithm; thus, we utilize the much simpler procedure of extrapolating the pressure at the boundary to cancel the source terms.  
We have checked the accuracy of our method by
keeping track of the the conservation of net magnetic flux and momentum
in the simulation domain.  The fractional error of these is no more than $10^{-12}$
and $10^{-6}$ respectively over thousands of dynamical times.

To seed the MTI, small amplitude perturbations are added to the vertical
velocity.  To seed a single mode, a perturbation of the form
\begin{equation}
\label{singlemode_v}
v_{z}(x) = v_{0}\sin\left(\frac{n\pi x}{L_{x}}\right),
\end{equation}
is used, where $n$ is the mode of the perturbation.
Note single mode perturbations are not smoothed at the
vertical boundaries.  To seed multiple modes, Gaussian white noise
velocity perturbations which are smoothed in the vertical direction using
$\sin (\pi z/L_z)$ are used.  The standard perturbation amplitude is
$v_{0}=10^{-4}$, which is much less than either the Alfv\'{e}n or sound
speed.

\section{Comparison to Linear Theory} \label{linear}

\subsection{Weak Field Limit} \label{linearweakfield}

As a test of our numerical algorithms, we have performed a series of
simulations using a grid of $100^{2}$ and single mode perturbations where
$\chi_C$ varies from $2.5\times 10^{-5}$ to $10^{-3}$ and $\chi_R = 0$
with a perturbation wavelength equal to the horizontal domain size.
The initial conditions contain a uniform horizontal field of uniform amplitude
$B_{0}/\sqrt{4\pi}=10^{-3}$ (net flux) and an initial velocity perturbation of
amplitude $v_{0}=2\times 10^{-4}$.  Adiabatic boundary conditions are
used on the vertical boundaries.  We measure the linear growth rate in
the simulation by extracting the horizontally-averaged amplitude of the
velocity perturbation seeded in equation \ref{singlemode_v} at different
times and applying the relationship,
\begin{equation}
\sigma=\frac{\ln(\bar{v}(t_{2})/\bar{v}(t_{1}))}{t_{2}-t_{1}}.
\label{gr_determination}
\end{equation}
The measured growth rates are plotted as crosses on the dispersion relation
for linear modes in the weak field limit shown in Figure 1.

The measured growth rates from the simulations are in excellent
agreement with the predictions of linear theory.  Given that the
Brunt-V\"ais\"al\"a frequency and temperature vary vertically over
the simulation domain, the initialized abscissa is actually a range,
the midpoint of which is plotted.  For example, one simulation has a
range of $\left(\chi'Tk^{2}/N\right) \in [1.019,1.072]$ which gives a
theoretical growth rate of $\sigma_{\mathrm{theory}}\in [0.279, 0.289]$.
Analysis of the linear phase of the simulation gives a growth rate
as $\sigma_{\mathrm{simulation}}\in [0.279, 0.295]$, a relatively
small error.  Due to the difficulty of defining a systematic error
convention for this case, error bars are not shown, and the average
values are plotted.  We conclude our numerical algorithms for anistropic
conduction accurately capture the MTI.

\subsection{Dependence on Magnetic Field Strength} \label{linearmag}

Like the MRI, the magnetothermal instability can be stabilized at large $k$
by tension.  Equation \ref{MTI_instability_criterion1} shows
the dependence of the MTI instability criterion on the magnetic field
(implicitly through the Alfv\'{e}n velocity).  We define a stability
parameter, $\eta$, as
\begin{equation}
\eta = \frac{k^{2}v_{A}^2}{\sigma^{2}_{\mathrm{max}}},
\label{etadef}
\end{equation}
where the maximum growth rate, $\sigma_{\mathrm{max}}$, is given by
\begin{equation}
\sigma^{2}_{\mathrm{max}} = \frac{\chi'_{C}}{\rho \chi'}\frac{\partial P}{\partial z}\frac{\partial \ln T}{\partial z}.
\label{maxgrowthrate}
\end{equation}

Figure 2 shows the effect of varying the magnetic field on the growth
rate holding all other parameters fixed.  We vary $B_{0}/\sqrt{4\pi}$
from $10^{-4}$ to 0.03 while the conductivity is held fixed
at $\chi_{C}=10^{-4}$ and $\chi_{R}=0$.  All other parameters are the
same as specified in the previous section.  As the magnetic field (and
stability parameter) approach zero, the growth rate asymptotes to its
maximum value given by the weak field dispersion relation, equation
\ref{dispersionrelation}.  As the magnetic field increases in strength
the growth rate decreases until $\eta=1$, at which point the instability
is entirely quenched.  The curve in Figure 2 is approximated roughly
by the fit 
\begin{equation}
\sigma(\eta) = \left\{ \begin{array}{ll}
(1-\eta)\sigma_{\mathrm{max}} & \eta \leq 1, \\
0 & \eta > 1.
\end{array} \right.
\label{eta_fit}
\end{equation}

\subsection{Effect of Isotropic Conduction} \label{lineariso}

Isotropic conduction can damp the linear growth rates of the
MTI, thus limiting its effect to regions where anisotropic conduction
dominates.
The anisotropy factor, $Q$, is defined as
\begin{equation}
\label{qdef}
Q\equiv \frac{\chi_{C}'}{\chi'}=\frac{\chi_{C}}{\chi_{C}+\chi_{R}},
\end{equation}
Equation \ref{dispersionrelation} can be easily rewritten in terms of $Q$ as
\begin{equation}
\left(\frac{\sigma}{N}\right)^{3} + \frac{1}{\gamma}\left(\frac{\sigma}{N}\right)^{2}\left(\frac{\chi' T k^2}{N}\right) +\left(\frac{\sigma}{N}\right) + \frac{d \ln T}{d \ln S}\left(\frac{\chi' T k^2}{N}\right)Q = 0. 
\label{dispersionrelation2}
\end{equation}

To study the effect of isotropic conduction, we hold the total
value of the conductivity, $\chi_{\mathrm{tot}}=\chi_{C}+\chi_{R}$
fixed.  Alternatively, this can be understood as choosing a fixed
non-dimensionalized wavenumber in the dispersion relation.  We then
vary the anisotropy factor, $Q$.  This linear study uses conducting
boundary conditions at a resolution of $100^2$.  Initial conditions are
a magnetic field of $B_{0}/\sqrt{4\pi}=5\times 10^{-4}$ yielding a net magnetic flux
and a single-mode perturbation.   Figure 3 shows the effect of isotropic
conductivity on the growth rate for a fixed $\chi_{\mathrm{tot}}=10^{-4}$.
The solid line in this plot comes from the solution of equation \ref{dispersionrelation}.

As the conduction becomes purely anisotropic, the growth rate
asymptotically approaches the weak-field value calculated previously.
When isotropic conduction dominates, the instability is totally
suppressed.  The measured growth rates are slightly over-estimated
compared to linear theory for $Q\sim 1$.  If one examines the value
$Q=0.01$, weak-field theory predicts a finite growth-rate; however, the
simulation shows zero growth.  In this case, the two terms of Equation
\ref{MTI_instability_criterion1} are of similar order and the weak-field
treatment breaks down.

An important result to note is that a small amount of isotropic
conduction, $Q = 0.99$, does not change the growth rate significantly from
no isotropic conduction, $Q = 1.0$.  This fact implies that the
anisotropic conduction routine does not need to preserve anisotropy to
as many orders of magnitude as is needed for tokamak plasma simulations.
We show in the appendix that our algorithm results in a ratio of parallel to
perpendicular conductivity of $> 10^3$ for arbitrary orientations of the
field with respect to the grid.  Thus, even though the effective $Q$
in our simulations is not as close to one as in a real plasma, it is
close enough, \textit{e.g.} $Q_{\rm eff} > 0.99$, that the resulting dynamics should
be unaffected.

\section{Non-Linear Regime of the Adiabatic Boundary Condition} \label{nonlinear}

Table 1 is a list of all runs using adiabatic boundary conditions
discussed in this paper. N1-N2 are characteristic non-linear studies
whose properties are displayed in this paper. Runs A1-A5 explore the
effect of initial magnetic field strength on saturation. Of course,
many more combinations of these parameters can and have been simulated.
An illustrative subset is presented here.

\subsection{Uniform Magnetic Field} \label{nonlinearfloat}

As a fiducial model, we consider a single-mode perturbation with a purely
anisotropic conductivity $\chi_{C} = 10^{-4}$ and a uniform magnetic field
strength $B_{0}/\sqrt{4\pi}$
of $5.0 \times 10^{-4}$  labeled Run N1 in Table 1.  Figure 4
shows plots of the magnetic field lines at various times.  The upper right plot shows the imprint of the sinusoidal perturbation on the magnetic
field quite well.  The lower left plot shows the significant non-linear
structure that has evolved.  Finally, the lower right plot shows the field
lines as they execute damped oscillation to a stable, albeit complicated,
saturated state.  This case has an initial total magnetic flux threading
the domain, so the magnetic field strength does not decrease with time,
and the magnetic flux remains constant to round-off error.

Figure 5 displays the components of the magnetic field energy,
$B_{x}^{2}/8\pi$ and $B_{y}^{2}/8\pi$, and of the kinetic energy density,
$\left(KE\right)_{x}=\rho V_{x}^{2}/2$ and $\left(KE\right)_{y}=\rho
V_{y}^{2}/2$, averaged over the computational domain.  The magnetic
energy associated with the horizontal component of the field is
amplified from the initial field strength and a vertical magnetic field
is created---both reaching a steady state value.  The kinetic energy
initially increases but asymptotes to zero, as the motion is only
facilitating the rearrangement of the global thermodynamic gradients.
The time is non-dimensionalized by the average adiabatic sound crossing
time of the domain which is $\bar{\tau_{s}}\approx 7.8\times 10^{-2}$.
For reference, the average Alfv\'{e}n crossing time for these parameters,
$\bar{\tau_{A}}\approx 193.4$, is much larger.  The conduction time,
$\tau_{C} \sim L_{x}^{2}/\chi_{C} = 100$, is also much larger.

Figure 6 shows the evolution of the horizontally averaged temperature
profile at various times.  Clearly, the profile evolves from a
monotonically decreasing temperature to a vertically isothermal state.
When the source of free energy is exhausted, \textit{i.e.} the temperature
is isothermal, the instability shuts off and the system reaches its new
hydrostatic state.  Despite the increase in the magnetic energy during the
evolution of the instability, the magnetic pressure is still negligible.
The final equilibrium state is simply a hydrostatic isothermal atmosphere.

Thus, we find that for adiabatic boundary conditions, the nonlinear 
saturated state of the MTI is an isothermal temperature profile.
These results are
consistent with intuitive expectations.

\subsection{Non-Uniform Magnetic Field}

Our fiducial model Run N1 contains a uniform field and therefore a
net magnetic flux.  We have performed simulations of the nonlinear
evolution of models with a sinusoidally varying horizontal field with
no net flux, in order to investigate whether the final saturated state
is affected.

Run N2 is identical to the fiducial model, except the that it is
perturbed with
Gaussian white noise to seed multiple modes, and that it
contains zero net initial
magnetic flux.  Figure 7 shows the initial, linear, non-linear, and
saturated states of the magnetic field at various times.  The magnetic
field structures here are much more complex than in the single-mode case.
Perhaps the most significant difference is the effect of magnetic reconnection.

Figure 8 plots the time evolution of horizontally-averaged magnetic and kinetic energies for Run N2.  A significant contrast to the previous case is the decay
of the magnetic field energy with time as a result of the anti-dynamo
theorem \citep{cowl34} and reconnection.

The time evolution of the vertical temperature profile is within $0.5\%$
of the previous case, and therefore is not shown.  Through comparison of
Run N1 and N2, the saturated hydrodynamic state seems to be independent of
the single/multi-mode and zero/nonzero magnetic flux initial conditions.
In addition, the saturated temperature profile does not depend on the
initial magnetic field strength as long as $k^{2}v_{A}^{2}\ll 1$.

\subsection{Dependence of the Saturated State on Resolution and Box Size}

In previous studies of the MRI \citep{hgb95} the effect of box size
and resolution have been quantified.  It is worth noting that our two
dimensional simulations of the magnetothermal instability reflect a
similar trend with respect to these simulation parameters.  First, we
define the magnetic field energy density amplification factor relative
to the initial state as
\begin{equation}
\label{deltaeqn}
\Delta=\frac{B^{2}(t=t_{sat})}{B^{2}(t=0)},
\end{equation}
where, $t_{sat}$, is the time at saturation.  As was seen with the
MRI, the saturation magnetic energy increases only very slightly with
increasing resolution.  Secondly, saturation magnetic energy scales
roughly linearly or slightly greater than linearly with the size of the
simulation domain.  Of course, this measurement must be repeated in a
three dimensional calculation to verify these results, as the anti-dynamo
theorem changes the underlying physics significantly in 3D.

\subsection{Dependence of the Saturated State on the Field Strength}

As was shown previously in \S\ref{linearmag}, the linear growth rate
relative to maximum growth rate (eq. \ref{maxgrowthrate}) of the
MTI is strongly dependent upon the magnetic field strength.  In the case
where the conductivity is entirely Coulombic, $\chi'_{c}=\chi'$, the
stability parameter, $\eta$, is independent of the thermal conductivity.
In addition to the dependence of the linear growth rate on the stability
parameter, the saturated state is also dependent on the strength of the
initial magnetic field.  Runs A1-A5 in Table 1 explore this dependence by
varying the initial magnetic field strength. This dependence manifests
itself primarily in two quantifiable ways.  First, the saturation
magnetic field energy density, $\Delta$, increases monotonically
as the magnetic field strength decreases; thus, reducing the role
of magnetic tension in retarding the growth of the instability.
As the magnetic field strength increases, the instability is completely
stabilized and the amplification factor is reduced to $1$.  The same
general pattern is also demonstrated by the dependence of the growth
rate on the stability parameter as seen in Figure 2.  This similarity
is not likely to be accidental.

The second effect observed from the variation of the magnetic field
strength is the degree of saturation obtained in terms of an isothermal
temperature profile.  Figure 9 shows various saturated temperature
profiles taken at the same time but with different stability parameters.
Where magnetic tension dominates, the temperature profiles
have evolved less from the initial conditions.  As the influence of
the magnetic field decreases, the temperature profiles become more
isothermal.  The profiles of simulations with higher magnetic field
would \textit{not} continue to evolve if the simulation were given
more time, as the instability is simply turned off before its free
energy source is completely used.  Essentially, the MTI reduces the
thermal gradient, the source of free energy, and amplifies the magnetic
field (increases the Alfv\'{e}n velocity) until the two terms of Equation
\ref{MTI_instability_criterion1} balance.  A new, stable  macroscopic
thermodynamic equilibrium is then established.

\section{Non-Linear Regime of the Conducting Boundary Conditions} \label{nonlinearfixed}

With adiabatic boundary conditions, there is a limited amount of free
energy available to the system, and once saturation occurs, the motions
decay away and a new vertical equilibrium corresponding to an isothermal
atmosphere (if $\eta \ll 1$) is established.  With conducting boundary
conditions, however, an isothermal profile across the domain can never
be achieved.  Free energy to drive motions can always be tapped from
the boundaries; therefore, we might expect that sustained convection
and turbulence are possible in this case.  To test these expectations,
we have performed a series of simulations of the nonlinear evolution of
the MTI using conducting boundary conditions.

Table 2 is a list of all runs using conducting boundary conditions
discussed in this paper.  
As a fiducial model, we consider a simulation with zero initial
net magnetic flux, run N3, which is perturbed with multiple modes.
An analogous case (Run N2) was considered in
\S4.2 for the adiabatic boundary condition.  Figure 10
shows the evolution of the magnetic field at various times.  The linear
evolution is not shown, but is quite similar to the adiabatic, multimode case
(see Figure 7).  After the system reaches a quasi-equilibrium,
bubbles of magnetic field emanate from the vertical boundaries and
penetrate to the center of the atmosphere.  These bubbles are reminiscent
of nucleate boiling familiar from fluid mechanics \citep{id}.  One can
observe the characteristic Kelvin-Helmholtz rolls on these bubbles.
This boiling ``steady-state'' behavior is reached after approximately
2,000 sound-crossing times as reflected by the plot of the magnetic and
kinetic energies shown in Figure 11.  It is clear that after an initial
spike some kinetic energy remains in the system, driven by the heat flux
at the boundaries.

The system is driven towards a more isothermal state with large boundary
layers at the upper and lower surfaces of the simulation (Figure 12).
Resolution studies show that as the simulation resolution is increased
($128^{2}$ is shown) the boundary layers shrink, and the interior region
becomes closer to isothermal.  We conclude that the fluid motion observed
at late times is driven by the MTI in an unresolved thermal boundary
at the top and bottom edges of the domain.  The resolution dependence
of the boundary layer makes it difficult to determine the efficiency of
heat transport by the magnetothermal instability in this case.

In fact, the resulting vertical structure which emerges is not likely
to be applicable to most astrophysical systems.  The use of a solid
reflecting wall kept at a constant temperature leads to a thin thermal
boundary layer, and the structure of this layer determines the physics.
In a real atmosphere, we expect the size of this boundary layer to be
set by the size of the transition region from isotropic to anisotropic
conduction.  This will be explored further in \S5.2, where we use
isotropic conduction to introduce resolved layers near the boundaries which are
stably stratified with respect to the Balbus criterion.  This modification avoids
the formation of thermal boundary layers at the wall.

\subsection{Dependence of the Saturated State on Isotropic Conduction}

As was shown in \S\ref{lineariso}, the linear growth rate of the MTI
is damped by the addition of isotropic conduction.  This behavior is
similar in many respects to that of the strong magnetic field case.
The saturated state is also dependent upon the anisotropy parameter, $Q$.
Figure 13 shows the horizontally averaged temperature profiles at the
time of saturation as calculated from runs I1-I4.  As can be seen,
even a modest amount of isotropic conduction, $Q=0.9$, is able to
significantly reduce the saturation level of the instability.  When the
isotropic conduction is strong, $Q=0.1$, the temperature profile only
marginally deviates from the initial condition.  The driving term of
the instability is reduced by the anisotropic fraction, thus, causing
saturation after much less growth.  In addition, the saturation magnetic
energy density, $\Delta$, [Eq. (\ref{deltaeqn})] scales monotonically
with $Q$, asymptotically reaching its maximum value for purely anisotropic
conduction.

\subsection{Models with Convectively Stable layers} \label{constable}

A more realistic boundary condition is implemented by establishing an
atmosphere with layers at the top and bottom that are stable according to
the Balbus criterion.  This is accomplished by transitioning slowly from
pure anisotropic conduction to isotropic conduction at the boundaries
of the domain.  We have performed a simulation, Run N4, with
a domain
$L_{x}=0.1$, $L_{z}=0.2$ and a resolution of 100x250.  The average
horizontal sound crossing time is slightly higher than the previous
studies with $\bar{\tau_{s}}\approx 7.88 \times 10^{-2}$.  Multiple modes
are seeded with Gaussian noise smoothed towards the vertical boundaries.
The total conductivity is $\chi_{tot} = 10^{-4}$, while the isotropic
and anisotropic conduction coefficients $\chi_R$ and $\chi_C$ are chosen
such that the anisotropy parameter $Q$ is
\begin{equation}
Q(z) = \left\{ \begin{array}{ll}
\frac{5}{L_{z}}\left(z-\frac{L_{z}}{5}\right) &
\frac{L_{z}}{5} \le z \le \frac{2L_{z}}{5} \\
1 & \frac{2L_{z}}{5} \le z \le \frac{3L_{z}}{5} \\
1 - \frac{5}{L_{z}}\left(z-\frac{3L_{z}}{5}\right) &
\frac{3L_{z}}{5} \le z \le \frac{4L_{z}}{5} \\
0 & \textrm{otherwise.}
\end{array} \right.
\label{Q_profile}
\end{equation}
Since $\chi_{C}=Q(z)\chi_{tot}$ and $\chi_{R}=(1-Q(z))\chi_{tot}$, the
conduction is purely isotropic near the boundaries, purely anisotropic
in the center, and linearly interpolated in between.

The magnetic field strength is initialized to follow the same vertical
profile, so that $B_{x}(z)=B_{0}Q(z)$, with $B_{0}/\sqrt{4\pi} = 10^{-4}$.
With this definition, the magnetic field is non-zero only in those
regions that are at least slightly unstable by the Balbus criterion.
We can track the mixing of the stable and unstable layers by following
the location of the field, since in ideal MHD the field is ``frozen in''
to the fluid.

Figure 14 shows the $x$ and $z$ components of the magnetic
and kinetic energy averaged over the domain in Run N4.  This plot is not much
different than the previous driven case Run N3 (Figure 11), except the magnetic
field does not decay in time due to the presence of a net magnetic flux.
The background temperature profile only deviates slightly from the
initial condition, indicating a fairly steady state.  The most interesting
insights come from examining the field lines at several different times
as shown in Figure 15.  The far left and middle left plots show the unstable
central region in the linear and early non-linear phases.  By the middle right
plot, however, plumes driven by the
MTI have penetrated the convectively stable region, as evidenced by the
magnetic field plume with Kelvin-Helmholtz roll.  At a much later time,
when the system is essentially in steady state, much of the magnetic
field has been stored in the stable region through convective overshoot.
The phenomena of penetrative convection and overshooting has been
thorougly studied in the magnetoconvection of stellar atmospheres,
and it is likely this phenomenon is universal \citep{tob01, bru02}.
In three dimensions this behavior could have important implications for
a magnetic dynamo.

A further diagnostic is the comparison of the magnitude
of each component of the
horizontally-averaged heat fluxes at different heights.
In addition to the Coulombic (equation \ref{coulombic}) and radiative
(equation \ref{radiative}) heat fluxes, we define an advective heat flux,
\begin{equation}
\mathbf{Q_{adv}} \equiv \mathbf{V}\epsilon,
\label{advective}
\end{equation}
where, $\mathbf{V}$, is the fluid velocity, and $\epsilon=p/(\gamma
-1)$, is the internal energy.  Figure 16 plots the time evolution of
the vertical components of these three heat fluxes and the total heat
flux at the midplane, and at 80\% of the height of the domain in Run N4.
At the
former height only anisotropic conduction is present, while at the latter
only isotropic conduction is present.  At the midplane the oscillatory
advective heat flux is clearly dominant at any given instant in time;
however, averaged in time the advective heat flux contributes roughly
$2/3$ of the total heat flux.  The Coulombic flux, which is relatively
constant in time, contributes the remaining $1/3$.  This fact is evident
in Table 3, which lists the time- and horizontally-averaged total heat
fluxes in the vertical direction at three different vertical heights.

To consider the heat conduction efficiency of the
instability, we compare it to the expected heat flux across the simulation
domain for pure uniform isotropic conductivity, namely $Q_{0}\approx 3.33
\times 10^{-5}$.  The time-averaged heat conduction at the midplane is
$\left<Q_{tot, 50\%}\right>\approx 3.54 \times 10^{-5}$, which indicates
that the instability transports the entire applied heat flux.  Thus,
we conclude that the MTI efficiently transports
heat flux in this two-dimensional case.

\section{Summary and Potential Applications} \label{conclusion}

Given an atmosphere with $dS/dZ<0$ and $N^{2}>0$, one would expect it to
be stable to convection.  Yet, if the density is low enough such that
anisotropic heat conduction along magnetic field lines is important,
the atmosphere is in fact convectively {\em unstable}, and it will
establish a radically different hydrostatic equilibrium using the
free energy provided by the initial temperature gradient.  We refer
to the new stability criterion in this case as the Balbus criterion, and
the convective instability that results as the magnetothermal instability.

Using time-dependent MHD simulations, we have verified the linear
properties of the MTI predicted by Balbus \citep{bal00}.  The growth
rates and Balbus criterion measured from the simulations agree well
with linear theory.  In the nonlinear regime, the instability saturates
as an isothermal atmosphere.  Steady convective turbulence can be
driven if a fixed temperature difference is maintained across the
upper and lower boundaries.  In this case, the advective heat flux
carried by fluid motions is larger than the conductive heat flux by
a factor of two.  The amplitude of the conductivity is important only
in establishing the most unstable wavelength.  For purely anisotropic
conduction, the maximum growth rate is independent of the conductivity.
Thus, our simulations show that if the Balbus criterion is satisfied,
unstable modes will produce vigorous motions and a significant heat
flux, even if the conductivity is very small, and conduction times very long.
We have also used simulations to study the effects of magnetic tension and isotropic conduction on the saturated state; in both cases the MTI
saturates before an isothermal profile is established.

There are several important directions for future studies.  Three
dimensional simulations of driven, steady convection that explore the heat
fluxes and dynamo action in the MTI will be reported in a future paper.
In addition, applications of the MTI to specific astrophysical systems
are warranted.  Two potential applications are of obvious and immediate
interest.  The first is to explore whether the MTI can help explain the
nearly isothermal temperature profiles observed in the outer regions of
x-ray emitting gas in clusters of galaxies \citep{fab94}.  In this case, anisotropic
ion viscosity (which has been ignored in this study) may also be important.
The second is to understand
the effect of the MTI on radiatively inefficient accretion flows.
In particular, there is much interest in the transport properties of turbulence driven by convection \citep{nmq98, spb99, nia00, qg00} versus the MRI \citep{sp01, hbs01, bh02, hb02} in such flows.  However, since for diffuse plasmas the appropriate stability criterion is not the H\o iland but rather the Balbus criterion, it is important to investigate how this changes the structure of the flow.

\acknowledgements
We thank Tom Gardiner for his contributions to the Athena code used here.
IP acknowledges current support from the Department of Energy
Computational Science Graduate Fellowship and previous support from the
Department of Defense National Defense Science and Engineering Graduate
Fellowship.  JS acknowledges support from Princeton University, and 
NSF grants AST-0413788 and AST-0413744.

\appendix

\section{Anisotropic Heat Conduction}
The heat conduction equation for Coulombic conduction is
\begin{equation}
\frac{3}{2} \frac{Pd\ln P\rho^{-5/3}}{dt} = -\mathbf{\nabla} \cdot \mathbf{Q} = \mathbf{\nabla} \cdot [\mathbf{\hat{b}}(\chi_{C} \mathbf{\hat{b}} \cdot \mathbf{\nabla} T)],
\label{app_heatconduction}
\end{equation}
where $\mathbf{\hat{b}}$ is a unit vector in the direction of the magnetic field.
In Cartesian coordinates, the right hand side may be written as
\begin{equation}
\chi_{C} \left\{\frac{\partial}{\partial x}\left[\hat{b}_x\left[\hat{b}_x\frac{\partial T}{\partial x} + \hat{b}_y\frac{\partial T}{\partial y}\right]\right]+\frac{\partial}{\partial y}\left[\hat{b}_y\left[\hat{b}_y\frac{\partial T}{\partial x} + \hat{b}_y\frac{\partial T}{\partial y}\right]\right]\right\}
\label{app_rhs},
\end{equation}
where $\hat{b_{x}}=B_{x}/[B_{x}^{2}+B_{y}^{2}]$ is the $x$-component of
the magnetic field normalized by the magnitude of $B$.  In order to
develop a finite-difference representation of equation \ref{app_rhs}, note that the
Athena code stores the magnetic field at cell faces and the temperature
at cell centers.  Thus, the second $x-$derivatives can be differenced as:
\begin{equation}
\frac{\partial}{\partial x}\left(\hat{b}_x^2\frac{\partial T}{\partial x}\right)=\frac{\hat{b}_{x,i+1/2,j}^2\left(T_{i+1,j}-T_{i,j}\right) - \hat{b}_{x,i-1/2,j}^2\left(T_{i,j}-T_{i-1,j}\right)}{\left(\Delta x\right)^{2}},
\end{equation}
and similarly for the second $y-$derivative.  The cross derivatives are
considerably more complicated, requiring a 3x3 difference molecule
to incorporate the change in the direction of the magnetic field across
a cell.  In addition, the mixed derivatives are written in a symmetrized
way, so that the heat transport is manifestly conservative.  For example,
\begin{eqnarray}
\lefteqn{ \frac{\partial}{\partial x}\left(\hat{b}_x\hat{b}_y\frac{\partial T}{\partial y}\right)=\frac{\hat{b}_{x,i+1/2,j}\hat{b}_{y,i+1/2,j}\left[\left(T_{i+1,j+1}-T_{i+1,j-1}\right)+\left(T_{i,j+1}-T_{i,j-1}\right)\right]}{4\Delta x\Delta y} } \nonumber\\
& & -\frac{\hat{b}_{x,i-1/2,j}\hat{b}_{y,i-1/2,j}\left[\left(T_{i-1,j+1}-T_{i-1,j-1}\right)+\left(T_{i,j+1}-T_{i,j-1}\right)\right]}{4\Delta x\Delta y}.
\end{eqnarray}
The difference formula are implemented in an external module to the MHD
integrator in Athena.  The two are combined using operator splitting---that is 
equation A1 is updated after a full timestep of the MHD equations.
Sub-cycling is used when the stability limit for the
timestep used in the conduction module is smaller than the timestep resulting
from the MHD algorithm.

To test the accuracy of the diffusion modeule, we have computed the
diffusion of a Gaussian profile of temperature in a medium with uniform
thermal conductivities.  In our tests with straight magnetic field lines
at an arbitrary angle to the grid the above finite differencing yields
fractional L2 error norms that are $O(10^{-8})$, independent of angle.

More complex tests involve anisotropic conduction along curved
magnetic field lines.  One of the more challenging test problems we
devised is the conduction of heat along circular magnetic field lines.
Specifically the problem is defined on a domain spanning $\{x,z\} \in
\{(-1,-1)...(1,1)\}$.  A circular heat pulse is initialized in a region
of an annulus defined as
\begin{equation}
T(r,\theta) = \left\{ \begin{array}{ll}
T^{*} & \textrm{if} \ (0.5\le r \le 0.7) \textrm{ and } 
(\frac{11\pi}{12}\le \theta\le\frac{13\pi}{12})\\
T_{0} & \textrm{otherwise;}
\end{array} \right.
\end{equation}
where in our simulation $T^{*}=12.0$, $T_{0}=10.0$, and
$\chi_{\parallel}=0.01$.  Figure 17 shows the quantitative convergence
of the L1 error.  At low resolution a large fraction of the
error arises simply from the discretization of a curvilinear problem to
a Cartesian domain.

From conservation of energy we can estimate the effective cross-field
diffusion coefficient.  Recall Fourier's law of conduction for a heat
flux, $q''=k\partial T/\partial x$ in units of $ [W/m^2]$.  We can
calculate the total energy passing through an area as
\begin{equation}
\int_{0}^{t} \dot{E}(t')dt' = \int_{0}^{t}\chi_{\perp}\frac{\partial T(t')}
{\partial x}A_{\perp}dt',
\label{heatestimate}
\end{equation}
where $A_{\perp}$ is the cross-sectional area.   Take the 100x100
resolution case as a canonical example since it was used as our minimum resolution.  At time $t=200$ when the annulus is essentially isothermal,
the net heat that has leaked from the annulus is $\Delta E=0.041$.
By doing an order of magnitude estimate on equation \ref{heatestimate}
we find that $\chi_{\perp}\sim 4\times10^{-6}$\, implying that the
ratio $\chi_{\perp}/\chi_{\parallel}\sim 10^{-3}-10^{-4}$.  As we have
demonstrated previously, the instability is insensitive to isotropic
conduction of this order of magnitude.

\clearpage

\begin{deluxetable}{ccccccccc}
\tablecaption{Properties of Runs with Adiabatic Boundaries}
\tablehead{
\colhead{Run} &
\colhead{$\chi_{C}$} &
\colhead{$\chi_{R}$} &
\colhead{$B_{0}/\sqrt{4\pi}$} &
\colhead{$\eta$} &
\colhead{Flux\tablenotemark{a}} &
\colhead{Resolution} &
\colhead{Mode}}
\startdata
N1 & $1 \times 10^{-4}$ & 0.0 & $5.0 \times 10^{-4}$ & $2.82 \times 10^{-3}$ &
+ & $(100)^{2}$ & Single \\
N2 & $1 \times 10^{-4}$ & 0.0 & $5.0 \times 10^{-4}$ & $2.82 \times 10^{-3}$ &
- & $(128)^{2}$ & Multi \\
A1 & $1 \times 10^{-4}$ & 0.0 & $9.42 \times 10^{-2}$ & 1.0 &
+ & $(100)^{2}$ & Single \\
A2 & $1 \times 10^{-4}$ & 0.0 & $6.94 \times 10^{-2}$ & 0.543 & 
+ & $(100)^{2}$ & Single \\
A3 & $1 \times 10^{-4}$ & 0.0 & $4.47 \times 10^{-3}$ & 0.225 & 
+ & $(100)^{2}$ & Single \\
A4 & $1 \times 10^{-4}$ & 0.0 & $2.0 \times 10^{-3}$ & $4.51 \times 10^{-2}$ &
+ & $(100)^{2}$ & Single \\
A5 & $1 \times 10^{-4}$ & 0.0 & $9.42 \times 10^{-4}$ & $1.0 \times 10^{-2}$ &
+ & $(100)^{2}$ & Single \\
\enddata
\tablenotetext{a}{'+' indicates net magnetic flux threading the domain, '-' indicates zero net magnetic flux.}
\end{deluxetable}

\begin{deluxetable}{ccccccccc}
\tablecaption{Properties of Runs with Conducting Boundaries}
\tablehead{
\colhead{Run} &
\colhead{$\chi_{C}$} &
\colhead{$\chi_{R}$} &
\colhead{$B_{0}/\sqrt{4\pi}$} &
\colhead{$\eta$} &
\colhead{Flux} &
\colhead{Resolution} &
\colhead{Mode}}
\startdata
N3 & $1 \times 10^{-4}$ & 0.0 & $5.0 \times 10^{-4}$ & $2.82 \times 10^{-3}$ &
- & $(128)^{2}$ & Multi \\
I1 & $1\times 10^{-4}$ & 0.0 & $5 \times 10^{-4}$ & $2.82 \times 10^{-3}$ &
+ & $(100)^{2}$ & Single \\
I2 & $9\times 10^{-5}$ & $1\times 10^{-5}$ & $5 \times 10^{-4}$ 
& $3.13 \times 10^{-3}$ & + & $(100)^{2}$ & Single \\
I3 & $5\times 10^{-5}$ & $5\times 10^{-5}$ & $5 \times 10^{-4}$ 
& $5.63 \times 10^{-3}$ & + & $(100)^{2}$ & Single \\
I2 & $1\times 10^{-5}$ & $9\times 10^{-5}$ & $5 \times 10^{-4}$ 
& $2.82 \times 10^{-2}$ & + & $(100)^{2}$ & Single \\
\enddata
\end{deluxetable}

\begin{deluxetable}{ccccc}
\tablecaption{Time-Averaged Vertical Heat Fluxes in Run N4 (see \S5.2)}
\tablehead{
\colhead{Height} &
\colhead{$\left<Q_{C}\right>$} &
\colhead{$\left<Q_{R}\right>$} &
\colhead{$\left<Q_{Adv}\right>$} &
\colhead{$\left<Q_{tot}\right>$}}
\startdata
20\% & 0 & $3.41 \times 10^{-5}$ & $-6.23 \times 10^{-6}$ &
$2.79 \times 10^{-5}$ \\
50\% & $1.39 \times 10^{-5}$ & 0 & $2.15 \times 10^{-5}$ &
$3.54 \times 10^{-5}$ \\
80\% & 0 & $3.30 \times 10^{-5}$ & $-2.97 \times 10^{-7}$ &
$3.27 \times 10^{-5}$ \\
\enddata
\end{deluxetable}

\newpage

\begin{figure}
\epsscale{.80}
\plotone{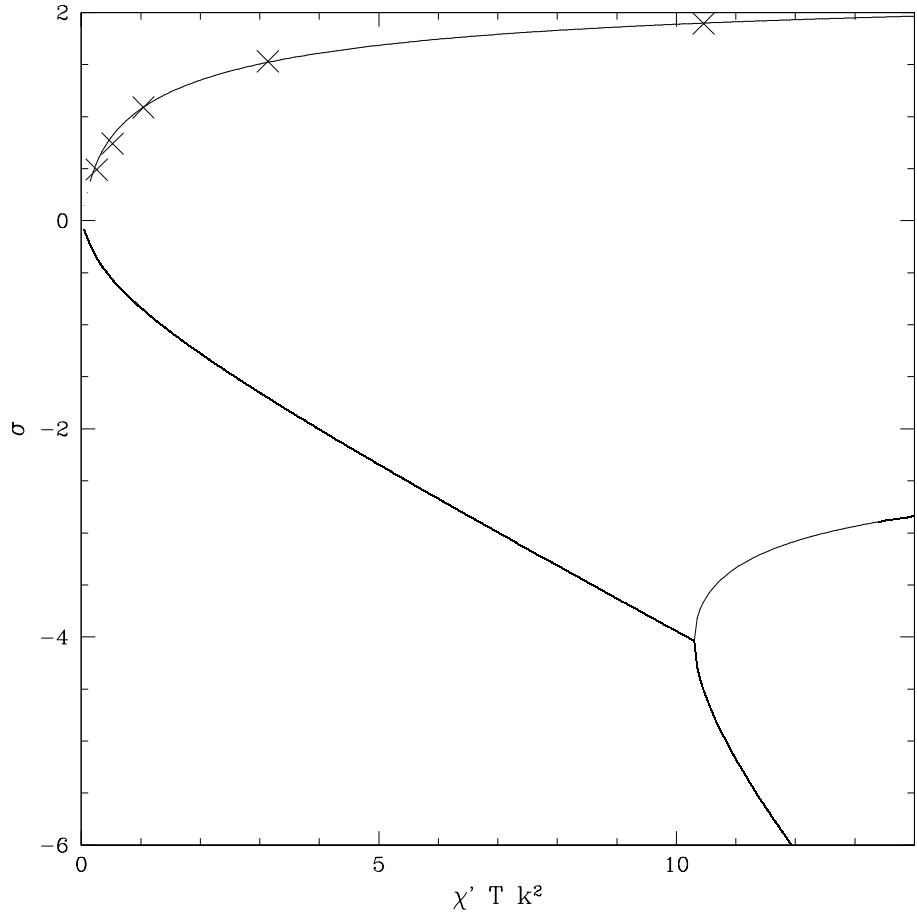}
\label{Plot_dispersion_relation}
\caption{Solutions of the MTI dispersion relation in the weak
field limit for an atmosphere with $d\ln T/d \ln S = -3.$  The axes
are normalized to the local Brunt-V\"{a}is\"{a}l\"{a} frequency, $N$.
The crosses are growth rates measured from simulations (see \S3.1).}
\end{figure}

\begin{figure}
\epsscale{.80}
\label{Plot_MTI_Alfven}
\plotone{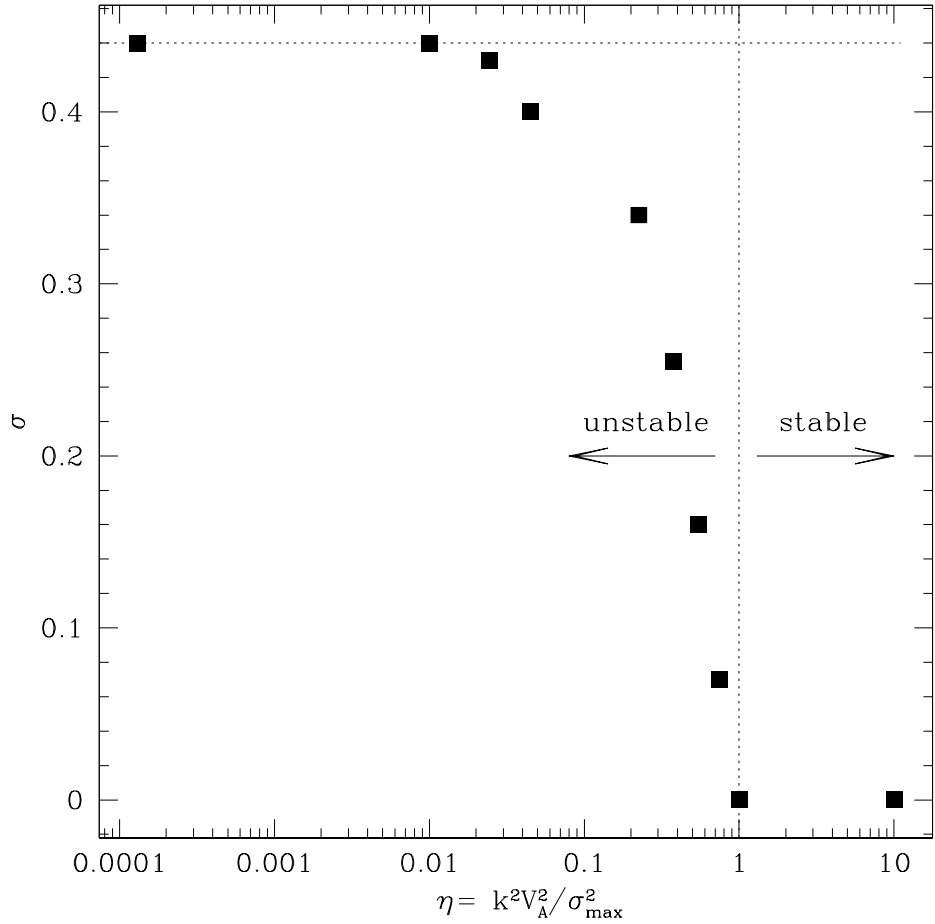}
\caption{The effect of magnetic tension on the MTI.  The growth rate
$\sigma$ decreases dramatically as magnetic tension
increases, \textit{i.e.} the stability parameter $\eta =
k^{2}V_{A}^{2}/\sigma^{2}_{\rm max}$ increases.  The transition from
stable to unstable is marked by a vertical dashed line, and the maximum
growth rate in the limit of weak magnetic field $\sigma_{\rm max}$
is marked by a horizontal dashed line.}
\end{figure}

\begin{figure}
\epsscale{0.80}
\plotone{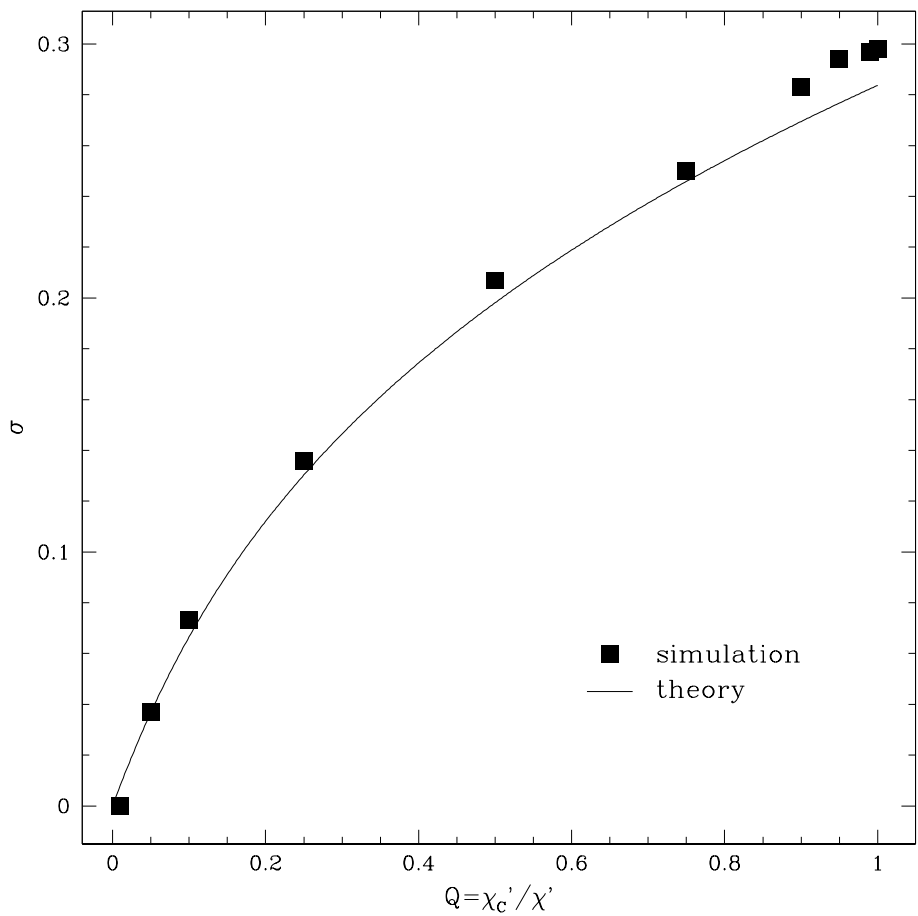}
\caption{The effect of isotropic conductivity.  Holding the total
conductivity $\chi_{\mathrm{tot}}$ and all other quantities fixed, the
anisotropic fraction, $Q$, is varied.  As $Q$ approaches 1.0, conduction
is purely anisotropic, and the growth rate reaches an asymptotic value.
As the conduction becomes more isotropic, \textit{i.e.} $Q$ approaches
zero, the instability is quenched.  The solid line marked 'theory' is
the unstable branch in the
solutions of equation 16}

\end{figure}

\begin{figure}
\epsscale{.80}
\plotone{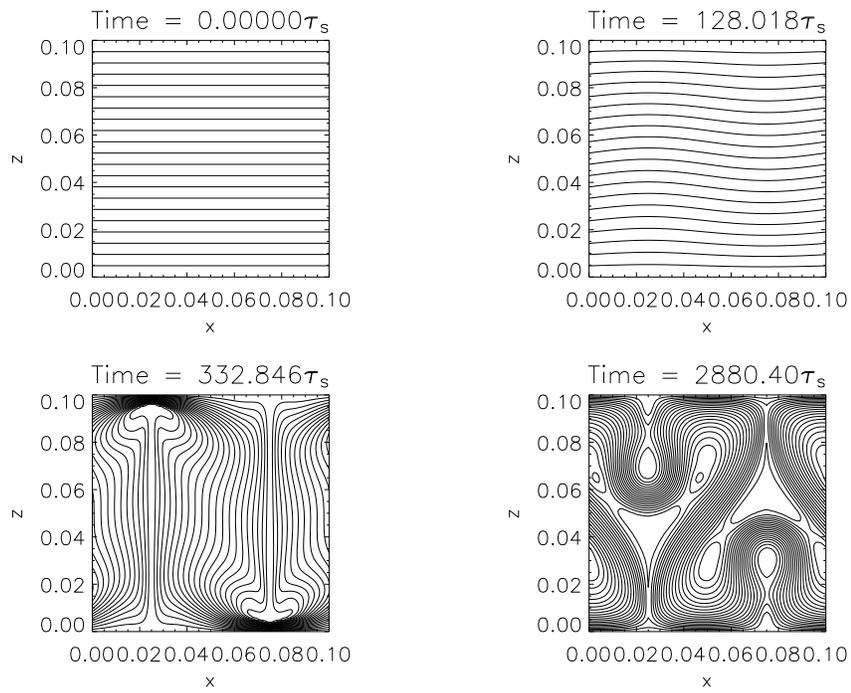}
\caption{Snapshots of the magnetic field lines in run N1 at various times
during the evolution of the instability. (upper left)
Inititial condition; (upper right) Linear phase; (lower left) Non-linear phase; (lower right) Saturated state.}

\end{figure}

\begin{figure}
\epsscale{0.80}
\plotone{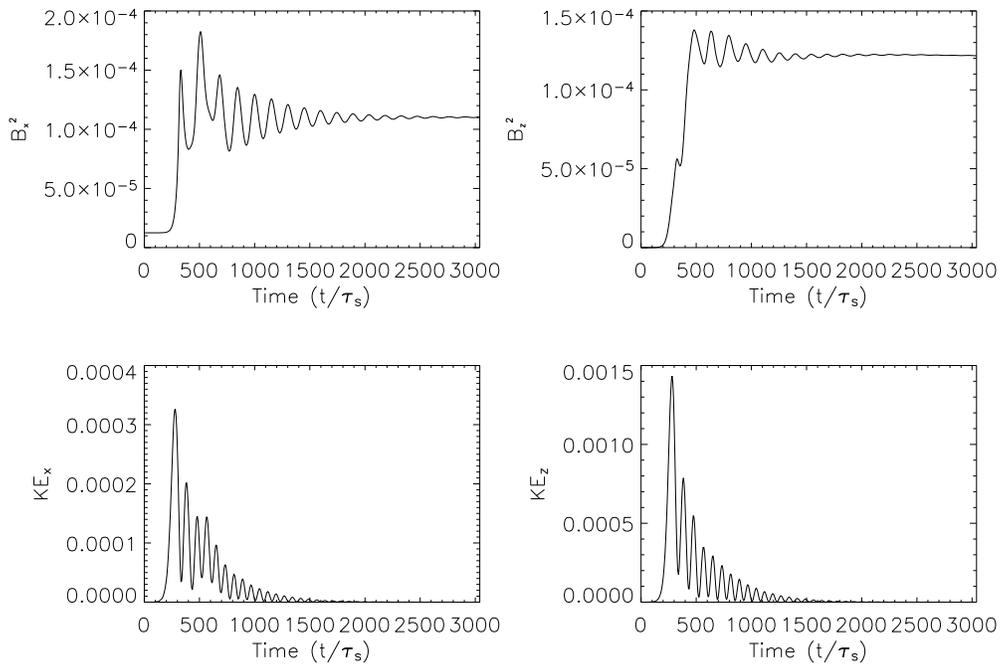}
\caption{Time evolution of the components of the volume averaged
magnetic energy and kinetic energy in run N1.  The time axis is
non-dimensionalized by the sound crossing time,
$\tau_{S}$.}
 \end{figure}

\begin{figure}
\epsscale{0.80}
\plotone{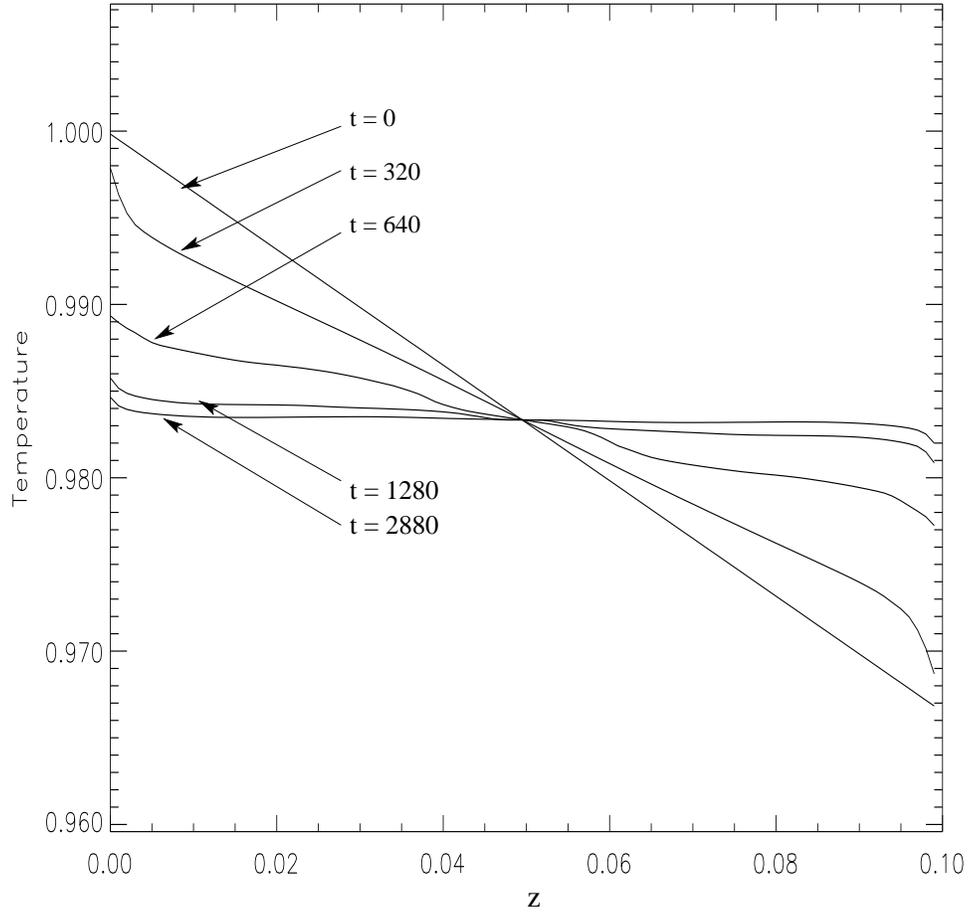}
\caption{Vertical profile of the horizontally-averaged temperature profile in
run N1 at various times.  The initial state
is a monotonically decreasing temperature profile with respect to height,
and the final state is isothermal.} \end{figure}

\begin{figure}
\epsscale{0.80}
\plotone{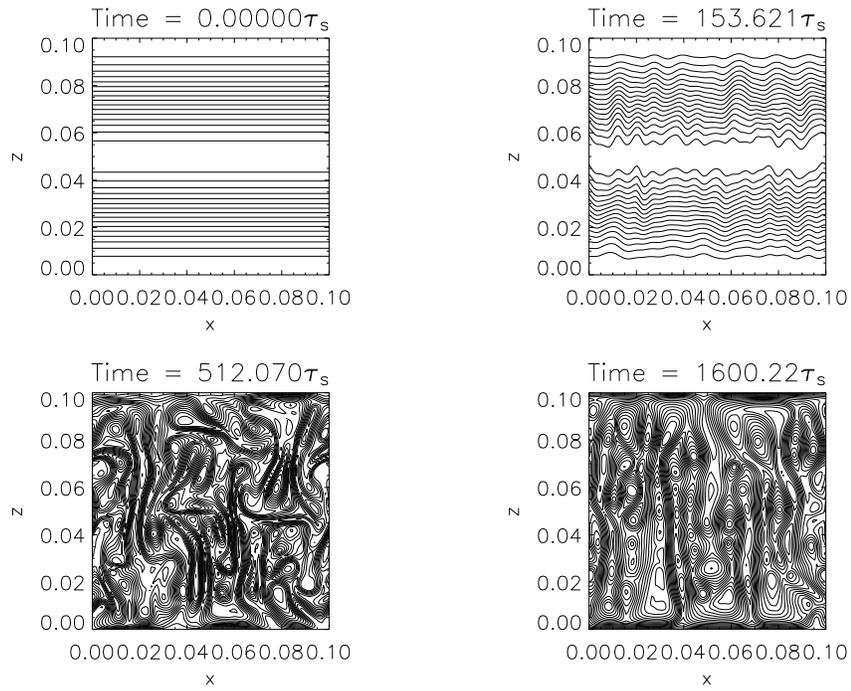}
\caption{Snapshots of the magnetic field lines at various times during the
evolution of the MTI in run N2. The structure is more complex than
the single-mode case.  (upper left) Initital condition;
(upper right) Linear phase; (lower left) Non-linear phase; (lower right) Saturated state.} \end{figure}

\begin{figure}
\epsscale{0.80}
\plotone{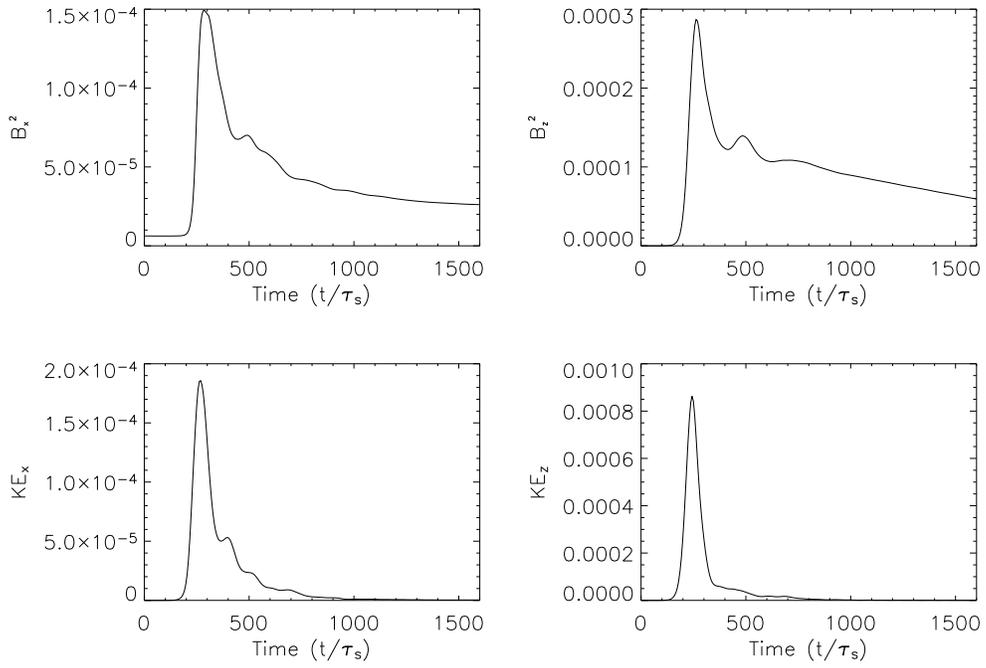}
\caption{Time evolution of the components of the volume averaged
magnetic energy and
the kinetic energy in run N2.  The magnetic field decays in time due
to the anti-dynamo theorem.} \end{figure}

\begin{figure}
\epsscale{0.80}
\plotone{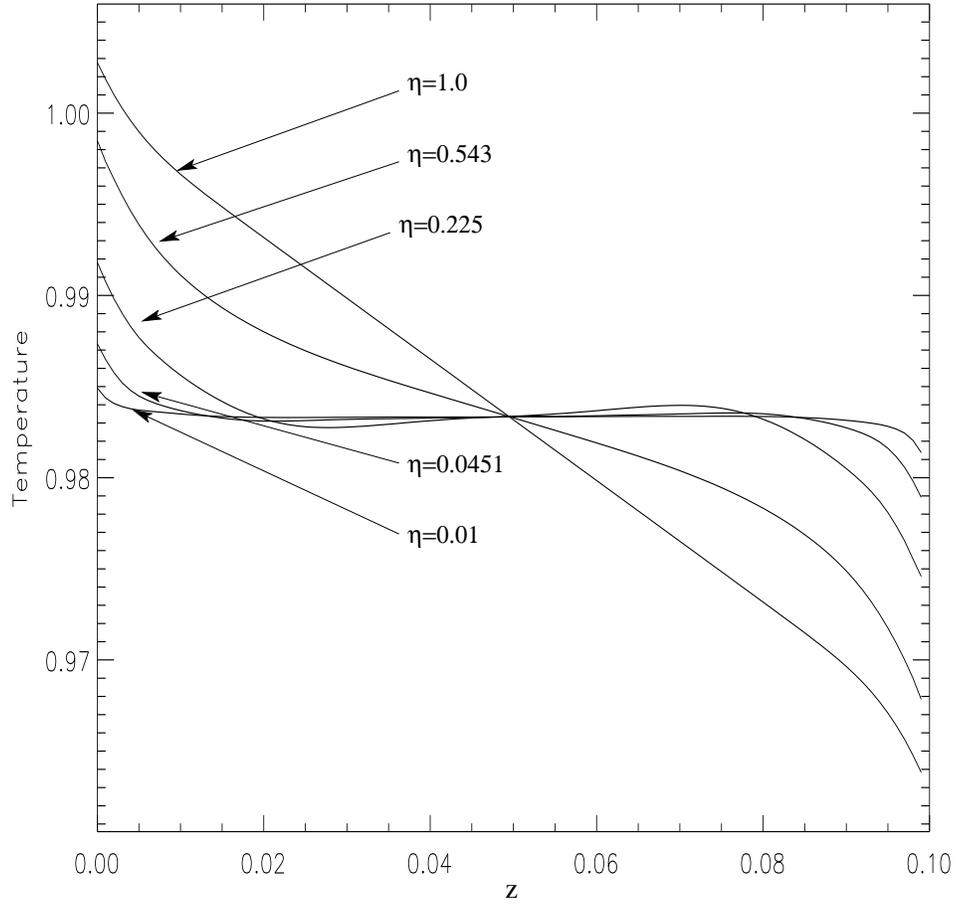}
\caption{Vertical profile of the horizontally averaged temperature at late
times in runs with various initial
magnetic field strengths (runs A1-A5).  As the stability parameter,
$\eta$, decreases, the saturated state asymptotically approaches an
isothermal profile.  States with sufficiently strong initial magnetic
fields \textit{never} saturate to an isothermal state.} \end{figure}

\begin{figure}
\epsscale{0.80}
\plotone{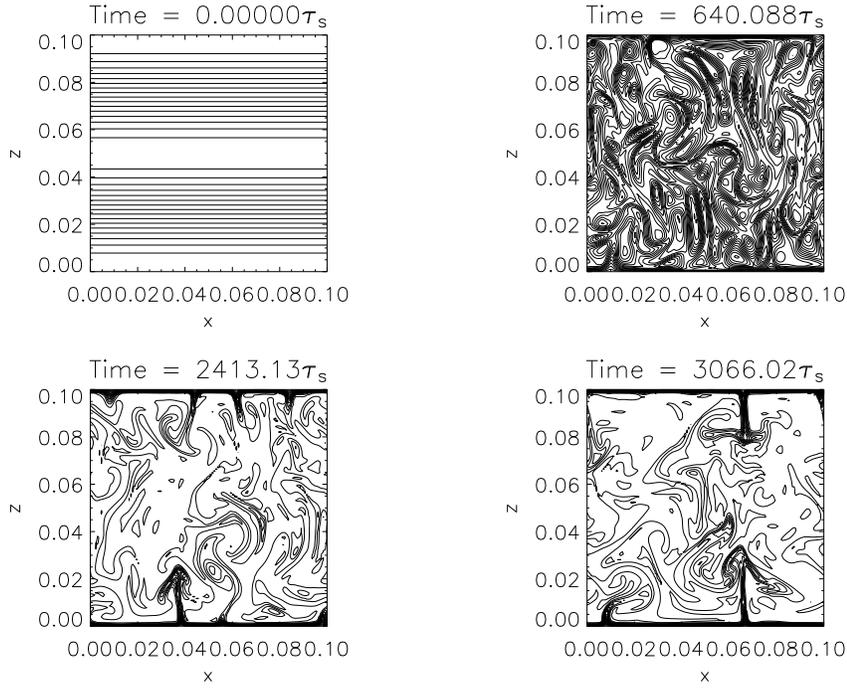}
\caption{Snapshots of the magnetic field lines at various times during
the evolution of run N3. (upper left)  The initial
magnetic field.  (uper right) Early non-linear growth is marked by highly complex
field geometry, reconnection, and generation of vorticity.  The bottom
plots show narrow plumes generated at the boundaries.
A thermal boundary
layer is evident at the upper and lower surfaces.} \end{figure}

\begin{figure}
\epsscale{0.80}
\plotone{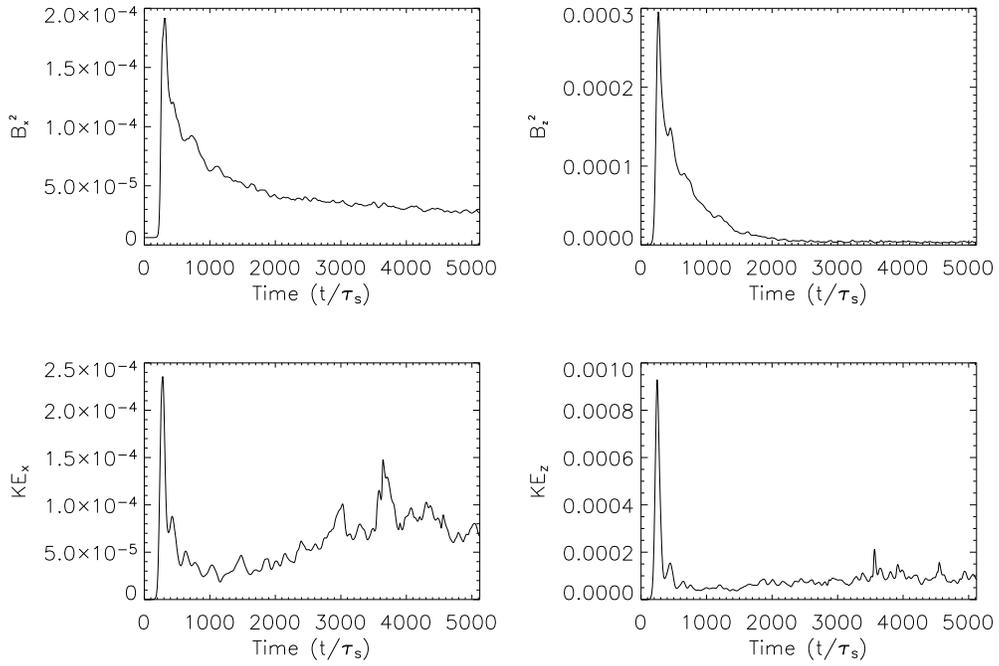}
\caption{Time evolution of the components of the volume averaged
magnetic energy and the kinetic energy in run N3.  The magnetic field
decays in time, but the kinetic energy continues to fluctuate due to 
the plumes generated at the boundaries.} \end{figure}

\begin{figure}
\epsscale{0.80}
\plotone{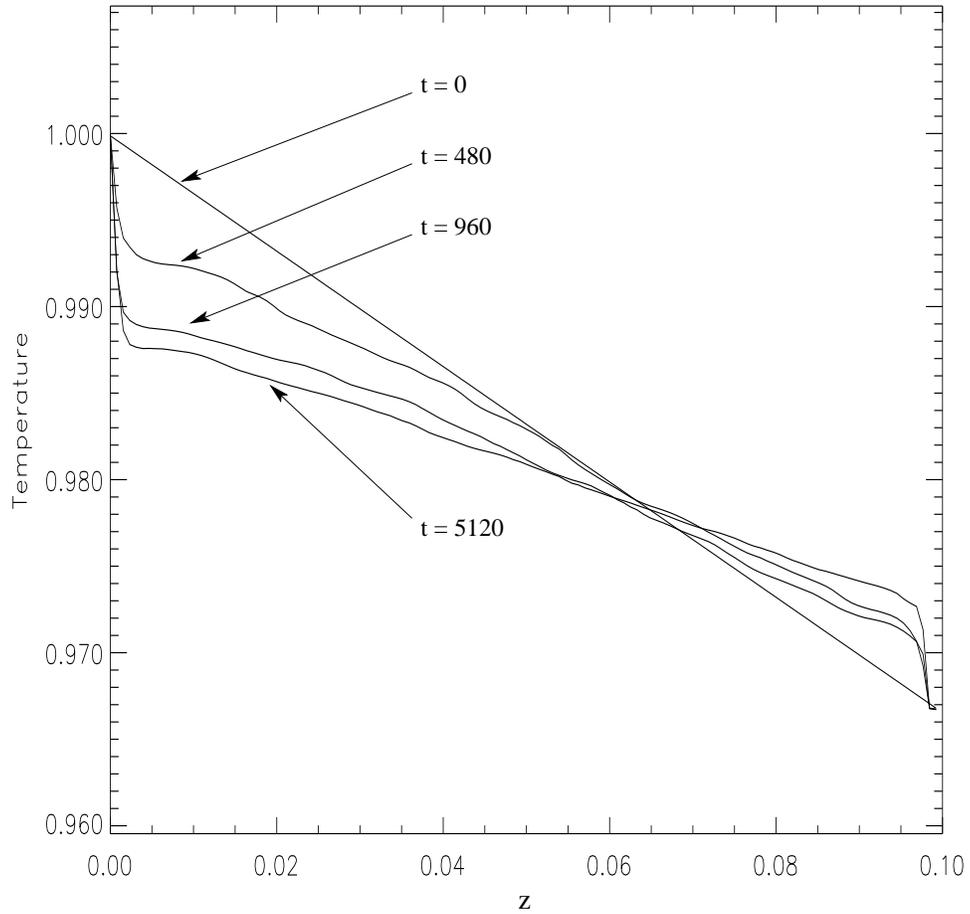}
\caption{Vertical profiles of the horizontally averaged temperature at various
times in Run N3.
A narrow thermal boundary layer develops, while
the interior saturates to a more isothermal profile than the initial
state. The width of the boundary layers is
resolution-dependent. } \end{figure}

\begin{figure}
\epsscale{0.80}
\plotone{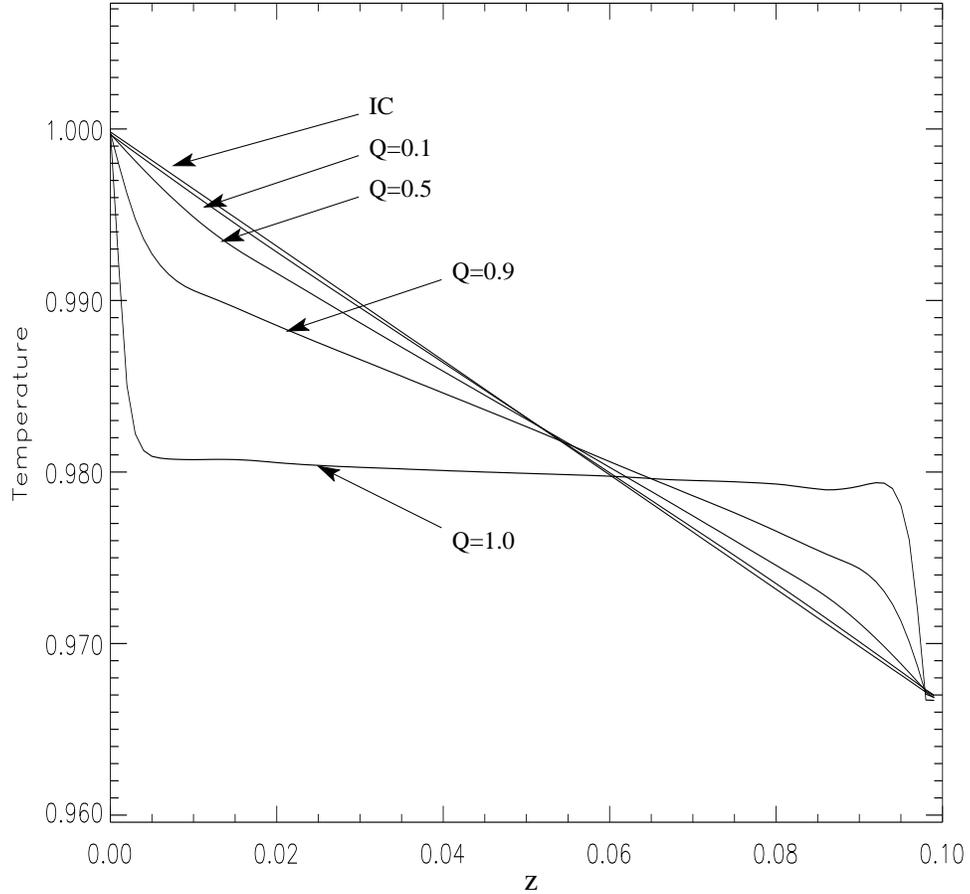}
\caption{Vertical profiles of the horizontally averaged temperature at
late times with various values of $Q=\chi_{C}'/\chi'$, the fraction of
anisotropic conductivity.  The initial condition is denoted `IC'.  It is
clear that even a modest amount of isotropic conductivity significantly
modifies the saturated state of the MTI.} \end{figure}

\begin{figure}
\epsscale{0.60}
\plotone{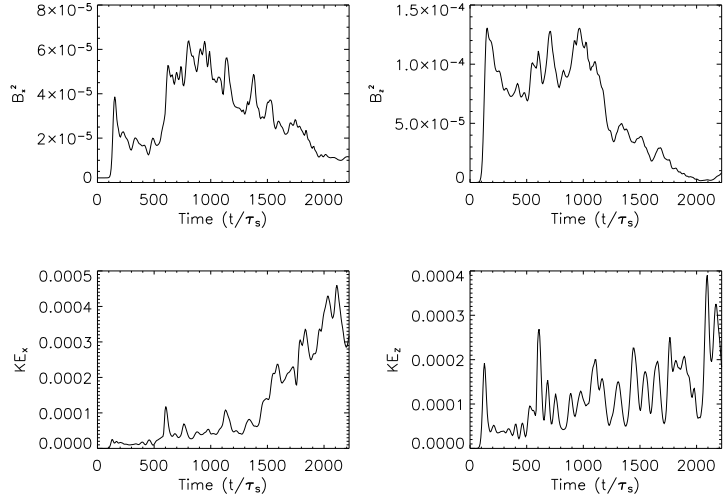}
\caption{Time evolution of the components of the volume averaged
magnetic and kinetic energy for Run N4.  }
\end{figure}

\clearpage

\begin{figure}
\epsscale{0.60}
\plotone{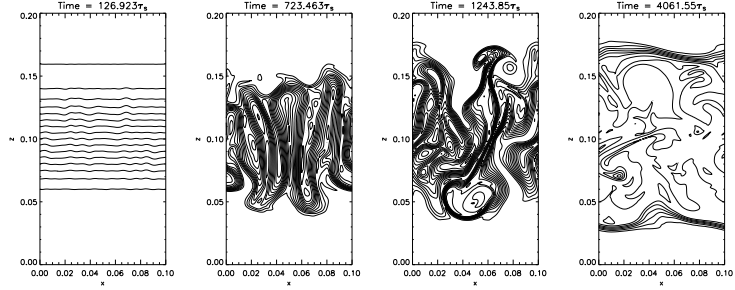}
\caption{Snapshots of the magnetic field in Run N4.
(far left) Early linear phase;
(middle left) early non-linear phase. (middle right) The MTI drives penetrative convection into the
stable layers, and at late times (far right) magnetic flux is pumped into the stable
layers.} \end{figure}

\begin{figure}
\epsscale{0.60}
\plotone{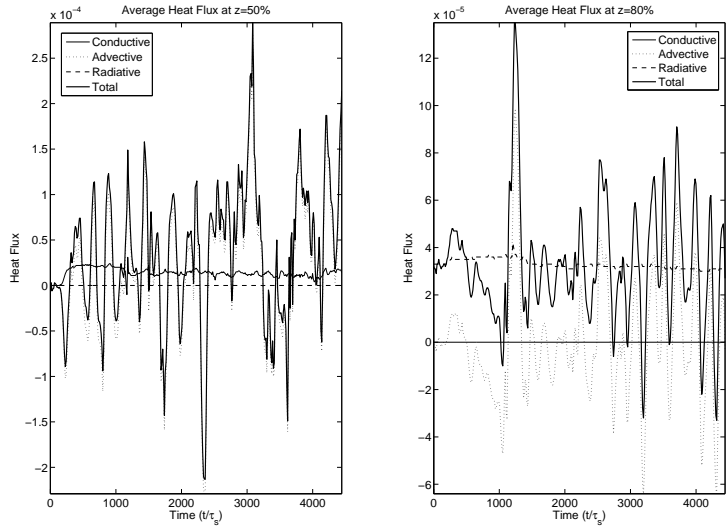}
\caption{Time evolution of the horizontally averaged heat flux at the
midplane and 80\% height of the simulation domain in Run N4.
The total heat flux
(thick solid line) is subdivided into Coulombic (thin solid line), radiative
(dashed lined), and advective (dotted line) components.  
The instantaneous advective heat flux generally is the dominant component of the total heat flux, particularly at the midplane.} \end{figure}

\begin{figure}
\epsscale{0.60}
\plotone{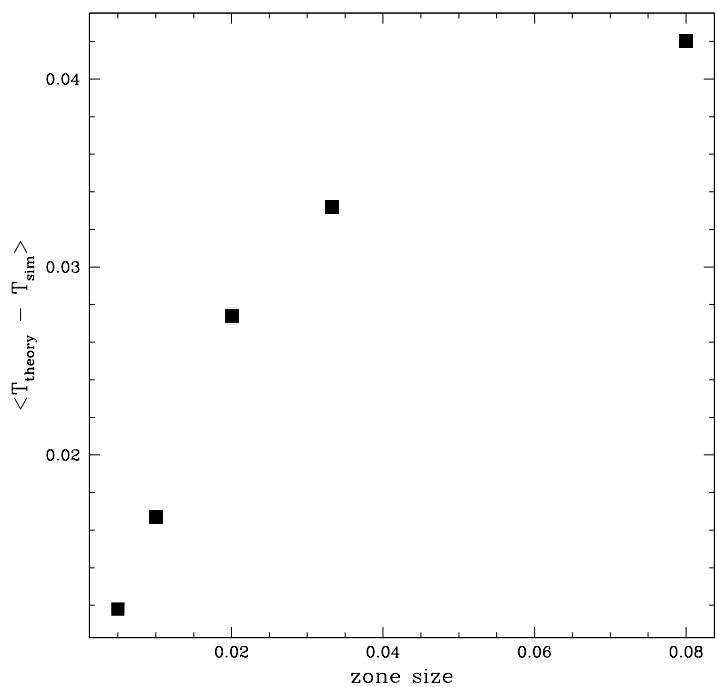}
\caption{The convergence of the average error with resolution for the circular
conduction test problem.}
\end{figure}


\begin{thebibliography}{}
\bibitem[Balbus \& Hawley (1998)] {bh98} Balbus, S. A., \& Hawley J. F. 1998, Rev. Mod. Phys., 70, 1
\bibitem[Balbus (2000)] {bal00} Balbus, S. A. 2000, \apj, 534, 420
\bibitem[Balbus (2001)] {bal01} Balbus, S. A. 2001, \apj, 562, 909
\bibitem[Balbus \& Hawley (2002)]{bh02} Balbus, S. A., Hawley, J. F. 2002, \apj, 573, 749
\bibitem[Balbus (2004)] {bal04} Balbus, S. A. 2004, \apj, 616, 857
\bibitem[Braginskii (1965)] {brag65} Braginskii, S. I. 1965,  in Reviews of Plasma Physics, Vol. 1, ed. M. A. Leontovich (New York: Consultants Bureau), 205
\bibitem[Brummell, Clune, \& Toomre (2002)] {bru02} Brummell, N.H., Clune, \& T.L., Toomre, J. 2002, \apj, 570, 825
\bibitem[Cowling (1934)] {cowl34} Cowling, T.G. 1934, \mnras, 94, 39
\bibitem[Fabian (1994)]{fab94} Fabian, A. C. 1994, ARA\&A, 32, 277
\bibitem[Gardiner \& Stone (2005)]{gs05} Gardiner, T. \& Stone, J. 2005, J. Comp. Phys., 205, 509
\bibitem[Hammett \& Perkins (1990)]{hp90} Hammett, G. W. \& Perkins, F. W. 1990, \prl, 64, 3019
\bibitem[Hawley, Gammie, \& Balbus (1995)]{hgb95} Hawley, J. F., Gammie, F., \& Balbus, S. A. 1995, \apj, 440, 742
\bibitem[Hawley, Balbus, \& Stone (2001)]{hbs01} Hawley, J. F., Balbus, S. A., \& Stone, J. M. 2001, \apj, 554, L49
\bibitem[Hawley \& Balbus (2002)]{hb02} Hawley, J. F. \& Balbus, S. A. 2002, \apj, 573, 738
\bibitem[Incropera \& DeWitt (1996)] {id} Incropera, F. \& DeWitt, D. 1996, Fundamentals of Heat and Mass Transfer (4th ed.; New York:Wiley)
\bibitem[Narayan, Igumenschev, \& Abramowicz (2000)] {nia00} Narayan, R., Igumenschev, I.V., \& Abramowicz, M.A. 2000, \apj, 539, 798
\bibitem[Narayan, Mahadevan, \& Quataert (1998)] {nmq98} Narayan, R., Mahadevan, R., Quataert, E. 1998, in Theory of Black Hole Accretion Disks, ed. M.A. Abramowicz, et al (Cambridge: Cambridge University Press), 148
\bibitem[Quataert, Hammett, \& Dorland (2002)] {qdh02} Quataert, E., Dorland, W., \& Hammett, G. W. 2002 \apj, 577, 524
\bibitem[Quataert \& Gruzinov (2000)] {qg00} Quataert, E., Gruzinov, A. 2000, \apj, 539, 809
\bibitem[Sharma, Hammett, \& Quataert (2003)] {sha03} Sharma, P., Hammett, G., Quataert, E. 2003, \apj, 596, 1121
\bibitem[Spitzer (1962)]{spitz62} Spitzer, L. 1962, Physics of Fully Ionized Gases (New York: Wiley)
\bibitem[Stone \& Pringle (2001)] {sp01} Stone, J. M. \& Pringle, J. E. P. 2001,\mnras, 322, 461
\bibitem[Stone, Pringle, \& Begelman (1999)]{spb99} Stone, J. M., Pringle, J. E. P. , \& Begelman, M. C. 1999, MNRAS, 310, 1002
\bibitem[Tobias, et al. (2001)] {tob01} Tobias, S. M., et al. 2001, \apj, 549, 1183
\bibitem[Zakamska \& Narayan (2003)]{zak03} Zakamska, N. L., \& Narayan, R. 2003, \apj, 582, 162

\end{thebibliography}
\end{document}